\documentclass[conference]{IEEEtran}
\makeatletter
\def\ps@headings{%
\def\@oddhead{\mbox{}\scriptsize\rightmark \hfil \thepage}%
\def\@evenhead{\scriptsize\thepage \hfil \leftmark\mbox{}}%
\def\@oddfoot{}%
\def\@evenfoot{}}
\makeatother
\usepackage{cite}
\usepackage{amsmath,amssymb,amsfonts}
\usepackage{graphicx}
\usepackage{textcomp}

\usepackage{subfigure}
\usepackage{algorithmic}
\usepackage{algorithmic}

\usepackage[ruled,vlined]{algorithm2e}
\usepackage{xcolor}
\usepackage{caption}
\usepackage{wrapfig,lipsum,booktabs}
\usepackage[export]{adjustbox}
\definecolor{red}{rgb}{2,0,0}
\usepackage{lipsum}
\usepackage{graphicx}
\ifCLASSOPTIONcompsoc
    \usepackage[caption=false, font=normalsize, labelfont=sf, textfont=sf]{subfig}
\else
\usepackage[caption=false, font=footnotesize]{subfig}
\fi
\def\BibTeX{{\rm B\kern-.05em{\sc i\kern-.025em b}\kern-.08em
    T\kern-.1667em\lower.7ex\hbox{E}\kern-.125emX}}

\IEEEoverridecommandlockouts    

\begin{document}
    
\title{A Resource Allocation Scheme for Energy Demand Management in 6G-enabled Smart Grid}

\author{
\IEEEauthorblockN{\textbf{Shafkat Islam}\IEEEauthorrefmark{1}, \textbf{Ioannis Zografopoulos}\IEEEauthorrefmark{2}, \textbf{Md Tamjid Hossain}\IEEEauthorrefmark{3},  \textbf{Shahriar Badsha}\IEEEauthorrefmark{4}, \textbf{Charalambos Konstantinou}\IEEEauthorrefmark{2}}

\IEEEauthorblockA{\IEEEauthorrefmark{1}Purdue University,
\IEEEauthorrefmark{2}CEMSE Division, King Abdullah University of Science and Technology (KAUST)\\
\IEEEauthorrefmark{3}University of Nevada, Reno,
\IEEEauthorrefmark{4}Bosch Engineering-North America\\
}

\IEEEauthorblockA{
E-mail: islam59@purdue.edu, mdtamjidh@nevada.unr.edu, shahriar.badsha@us.bosch.com \\
\{ioannis.zografopoulos, charalambos.konstantinou\}@kaust.edu.sa }

}

\IEEEaftertitletext{\vspace{-1.0\baselineskip}}

\maketitle

\begin{abstract}
Smart grid (SG) systems enhance grid resilience and efficient operation, leveraging the bidirectional flow of energy and information between generation facilities and prosumers.
For energy demand management (EDM), the SG network requires computing a large amount of data generated by massive Internet-of-things sensors and advanced metering infrastructure (AMI) with minimal latency. This paper proposes a deep reinforcement learning (DRL)-based resource allocation scheme in a 6G-enabled SG edge network to offload resource-consuming EDM computation to edge servers. 
Automatic resource provisioning is achieved by harnessing the computational capabilities of smart meters in the dynamic edge network.
To enforce DRL-assisted policies in dense 6G networks, the state information from multiple edge servers is required. However, adversaries can ``poison'' such information through false state injection (FSI) attacks, exhausting SG edge computing resources.
Toward addressing this issue, we investigate the impact of such FSI attacks with respect to abusive utilization of edge resources, and develop a lightweight FSI detection mechanism based on supervised classifiers. Simulation results demonstrate the efficacy of DRL in dynamic resource allocation, the impact of the FSI attacks, and the effectiveness of the detection technique. 
\end{abstract}


\section{Introduction} \label{s:Introduction}



Significant efforts have been made towards the digitization of the power grid. Smart grid (SG) functionalities enhance grid resilience and sustainability while enabling the efficient utilization of renewable and distributed resources, outgrowing its previous hierarchical and centralized nature (e.g., bulk coal, oil, etc. generation facilities). Leveraging concepts such as the Internet-of-Things (IoT), advanced metering infrastructure (AMI), demand-response schemes, and 6G connectivity, the power grid is transitioning from a unidirectional power architecture to a composite information and power network.

It is predicted that 530 GWs of distributed generation will be integrated into the power grid by 2024\footnote{www.fortunebusinessinsights.com}. The rapid rate of renewable and distributed penetration, the intermittency of such resources, and the timing constraints introduced by the energy market operation (e.g., demand-response schemes) require real-time administration. Centrally processing such information and coordinating with the SG end-nodes becomes time consuming (e.g., network latency) and resource-demanding (CPU resources, computer systems, information management, etc.). 
6G technologies and edge computing can provide solutions to the challenging problems of optimally predicting, allocating, and economically deploying available power system resources \cite{9237965, 9454590}. 
Leveraging the deployed computational power of embedded systems, controllers, AMI and IoT at the edge of the power grid, can alleviate the number of data that have to be centrally processed, and curtail unnecessary network data traffic (e.g., end-node to central server). Furthermore, using 6G, communication can be accelerated, allowing for more sophisticated schemes and ensuring resilient real-time operation.

The importance of 6G technologies to manage resources in heavily interconnected and reconfigurable SG deployments is the epicenter of many research works. For instance, the authors in \cite{reliability2021}, demonstrate the gravity of reliable wireless connections between edge devices in time-critical operations, as well as the potential consequences if such requirements (i.e., availability and reliability) are not met. To moderate the computational burden of dense SG networks, researchers have proposed offloading schemes leveraging the capabilities of edge nodes (ENs). In \cite{9449185}, a collaborative offloading strategy is proposed where after a central server has received all the expected requests, instead of evaluating them locally, the requests are fragmented and forwarded to EN devices for processing. In \cite{8761535}, task offloading scheme is formulated as a joint optimization problem, where the communication (AMI) and the computation (ENs) devices are cooperatively optimized. Q-learning approach has been explored to assist the optimal offloading decision making process \cite{9473888}. Using stochastic optimization, the offloading problem can be addressed, enabling the efficient operation of demand-response mechanism in decentralized SG topologies with renewable penetration \cite{9237149}.

The advanced interconnection and reconfigurability of SGs, the multitude of EN device types, and the transformation of the traditional power grid into a ``network of information'', introduces data security and reliability implications \cite{xenofontos2021consumer, hossain2021DeSMP, hossain2021privacy}. More specifically, the power grid utilities leverage the private and confidential operational data (generated by hundreds of interconnected sensors, intelligent electronic devices (IEDs), distributed energy resources (DERs), etc.) of the modern bidirectional SGs to detect potential cyber/physical anomalies (e.g., sudden power fluctuations), perform state estimation, fault or disturbance analysis, islanding detection, load modeling, forecasting, etc. \cite{zografopoulos2021detection, dileep2020survey, liu2020study}. Therefore, any stealthy cyberattacks by porting false state injection (FSI) into the system or by delaying and manipulating in-transit data may jeopardize the grid's stable operation \cite{rath2022closed, 9276446, zografopoulos2021cyber, zografopoulos2021security}.
In this work, to address both the computationally demanding resource allocation problem and the security requirements of densely interconnected SG systems, we propose the use of deep reinforcement learning (DRL) algorithms to identify optimal solutions to the time-sensitive energy demand management (EDM) problem. We utilize the computational power of ENs to deal with the advanced complexity of distributed optimal resource allocation, along with 6G technologies that can provide high-speed data throughput, minimal latency, and reliable communications that provide high quality of service in real-time applications. We also investigate the performance of DRL policy in the presence of FSI attacks and develop fine-grained FSI detection techniques based on supervised classifiers. \looseness=-1

\section{Problem Formulation} \label{s:problemFormulation}

\begin{figure}
\centerline{\includegraphics[width=0.48\textwidth]{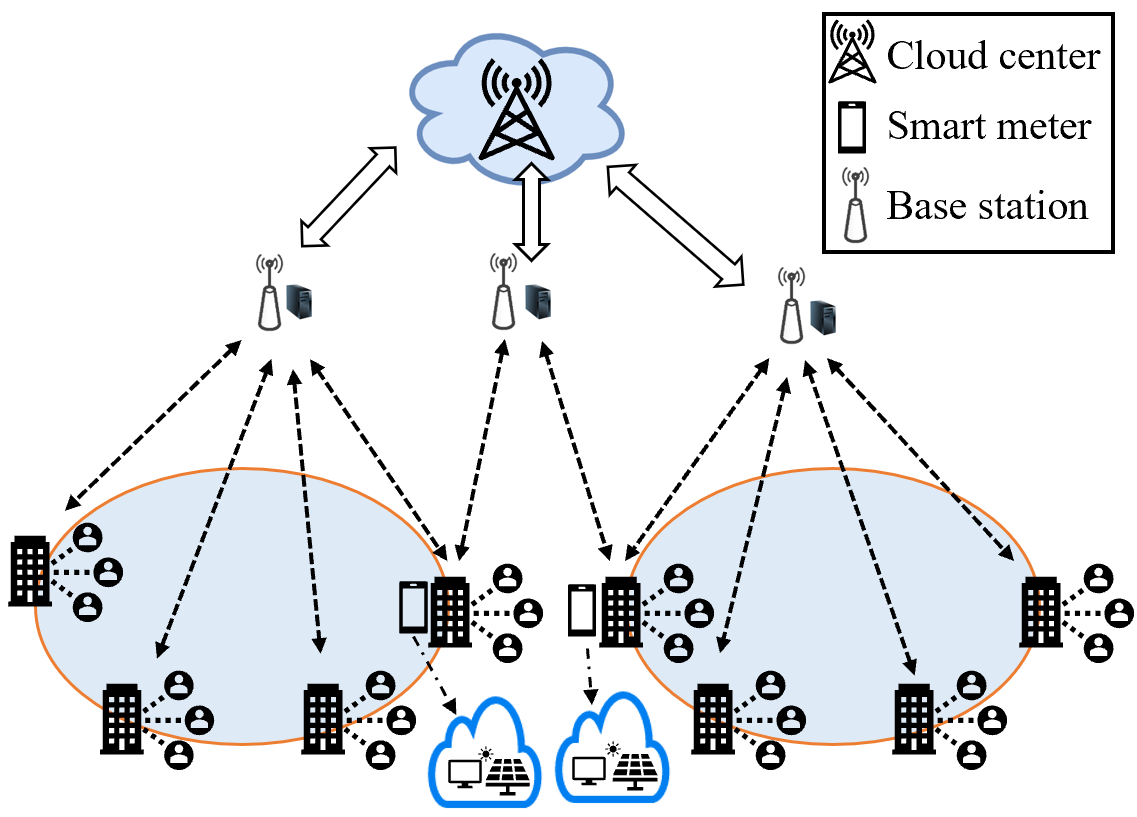}}
\caption{6G-enabled edge computing based SG network. }
\label{fig:system}
\end{figure}

\subsection{6G-enabled Edge Computing based Smart Grid Network}

Fig. \ref{fig:system} illustrates an edge-computing architecture for the residential SG system deployed in urban or suburban areas. In this system, we assume that renewable energy, i.e., solar photovoltaic panels, serves as an alternative energy source in addition to the traditional fuel-based system. The EDM scheme is installed on the edge servers, and we assume that each smart meter is under the coverage of multiple edge servers (since edge servers are densely deployed in 6G \cite{9237965}). We assume that the residential sector is divided into various geographic locations for the convenient dispatch of electric power.

\subsection{Energy Demand Management Task Workload}

The edge servers in this architecture host the computation-intensive optimal energy scheduling and price determination tasks. However, to execute these tasks efficiently, the edge server requires monitoring data and demand data from photovoltaic sensors and smart meters, respectively. To understand it clearly, we can consider subdividing the task into physical-layer and cyber-layer operations. 
\subsubsection{Physical-layer operation} In the physical layer of our proposed architecture, the generators, batteries, DERs (e.g., solar photovoltaic panels), etc. are producing and supplying energy through substations. They are also connected to certain controllers and loads to effectively balance energy demand and supply. We assume that several smart meters and sensors are physically deployed in the edge network and connected to loads and DERs through some wired or wireless mediums. They are primarily responsible for collecting generation and consumption data from the DERs and loads, respectively. Moreover, smart meters forward the collected data to the edge server for further processing in a periodic manner.
\subsubsection{Cyber-layer operation} The edge server periodically executes tasks and returns the energy scheduling policy to each smart meter for a certain period. However, in order to execute the task in each tenure, the execution server must collect data from all wireless monitoring sensors (monitoring renewable energy generation) and smart meters in a specific region. We denote each EDM task as, 
 $T_{EDM}=<ID_i, D_d, D_m, QoS, timestamp>$,
 which is a tuple of five different entities, namely, a unique task identification number $(ID_i)$, required demand $(D_d)$ and monitoring $(D_m)$ data, the quality-of-service ($QoS$) requirement (in terms of latency), and unique $timestamp$. Since this paper considers only a single type of task (energy scheduling and price estimation), the QoS requirement for each task is identical. 
\subsection{Computation Model}
We consider that the EDM task can only be executed on edge servers and that smart meters do not possess sufficient computing power to execute EDM tasks. This paper assumes that each edge server possesses identical computation power called CPU cycles/s $(f_i)$. Therefore, the execution time to perform the $j^{th}$ EDM task on the $i^{th}$ server can be defined by the following equation, as    $t_{EDM}^j=\frac{D_s}{r_i}+\frac{C_s}{f_i}$
\noindent where $D_s$ represents the sum of data (monitoring, $D_m$ and demand, $D_d$) required for the execution of the $j^{th}$ task, $r_i$ is the average transmission data rate, and $C_s$ is the total computation requirement (summation for processing $D_m$ and $D_d$). However, we only consider transmission energy cost and ignore the edge server task execution energy cost since the server usually possesses bulk energy for operation. Hence, we define the energy consumption for the $j^{th}$ task execution as, 
$E^j_{EDM}=\frac{P_i D_s}{r_i}$, where $P_i$ denotes the average transmission power\cite{9237965}.

\subsection{Communication Model}

We consider that the SG network is connected to a mobile wireless cellular network, and a high-speed optical fiber connection is used between the base station and the edge server \cite{9237965}. Thus, we ignore the latency between the server and the base station. As a result, the transmission rate between the smart meter and the base station can be defined as, 
    $r_i=\frac{B_k}{N}log_2(1+\frac{P_i C_{ik}}
    {\omega \sum_{l=1;l\neq i}^{N} P_i C_{ik}})$\cite{9237965},
where $B_k$ denotes the bandwidth of the $k^{th}$ edge server, $C_{ik}$ denotes the channel gain between the $i^{th}$ smart meter and $k^{th}$ edge server, and $\omega$ represents the background noise.

\subsection{Problem Definition}

The optimization objective of this work is to minimize the total execution overhead of smart meters tasks. We define  the task offloading problem in a 6G-enabled edge server \cite{9237965},  using the following equation,

\begin{equation}
\begin{aligned}
\min_{K} \quad & \xi\sum_{j=1}^{N}{ t_{EDM}+(1-\xi)\sum_{j=1}^{N} E_{EDM}}\\
\textrm{s.t.} \quad & C1 :  \sum_{k\in K} u_{ik}\geq 1; \forall{i \in S}\\
\quad & C2 :  \sum_{d_{ij}\in K} d_{ij} = 1; \forall{j \in T}\\
\quad & C3 :  d_{ij} \in \{0, 1\};\forall{d_{ij} \in K}\\
\quad & C4 : \lambda_k < \mu_k; \forall{k \in K}\\
\end{aligned}
\end{equation}

\noindent where $\xi$ serves as a balancing parameter and its value is within $[0,1]$, $S$ and $T$ denote the set of smart meters and tasks, respectively, $u_{ik}$ denotes the coverage of the $i^{th}$ smart meter by the $j^{th}$ server, $d_{ij}$ denotes the offloading decision for the $j^{th}$ task to the $i^{th}$ server, and $\lambda_k$ and $\mu_k$ denote the arrival rate and processing rate for the $k^{th}$ server. 

Since we assume that the EDM task is executed on public edge servers where the computation renters are smart meters and regular mobile users, the offloading optimization technique requires responsiveness in a dynamic edge environment. To address this constraint, we develop a DRL-based distributed offloading policy for smart meters. Moreover, we assume that edge servers utilize traditional cryptographic methods to preserve the data privacy of smart meters.
Thus, we focus only on FSI attacks targeting the offloading policy.

\subsection{Threat Model}

Our threat model considers state space manipulation attacks, i.e., FSI attacks aiming to compromise the dynamic offloading policy. Since in 6G, each task requires choosing a server for offloading from multiple servers (due to the overlap in coverage region), the offloading agent has to collect the current state of multiple covering edge servers. A malicious adversary can exploit this constraint and inject malicious edge server states that mislead the agent into making sub-optimal decisions, and as a result increase the offloading cost \cite{singh2021machine}. This type of attack can be launched by the edge servers themselves (insider threat) or by an external entity eavesdropping on the communication channel (outsider threat).

Besides increasing the offloading cost, the FSI attack can also bring devastating consequences to the inter- and intra-actions of the grid's physical layer equipment. For example, load balancing and generator synchronization are often used as two critical operations to meet demand and supply requirements. The grid controllers utilize the generator states and loads to determine the optimal next actions for effective synchronization and balancing. Now, an FSI attack on the edge server can manipulate the generator/load state, which in turn can mislead the agent to take incorrect or sub-optimal action. An improper generator synchronization or load balancing due to this deviation can introduce electrical and mechanical transients that may eventually lead to the failure of the entire grid. Therefore, an effective FSI detection technique is equally important for the safety of the physical layer equipment in a DRL-driven SG environment.

\section{Deep Reinforcement Learning-based offloading scheme}
\label{s:DRLscheme}

It is impractical to mathematically, and exhaustively, formulate the edge conditions of the EDM, therefore, we use a DRL-based offloading optimization technique.

\subsection{State Space} 

We define the state space of the agent as $S = \{D_d, D_m, C_d, C_m, DR_1, ....., DR_k, L_1, ...., L_k, QoS\}$, where $C_d$ and $C_m$ represent the processing requirements for demand and monitoring data, $DR_1, ..., DR_4$ and $L_1, ..., L_4$ denote the data rate and average queuing latency of the base stations, respectively, and $QoS$ represents the latency requirement of the EDM task. Under nominal circumstances, the values of each state feature can be continuous within specific ranges.

\subsection{Action Space} 

In DQN, the agent takes an action in each state based on the observation of the features in that state. Since the offloading DQN agent is assigned to determine the optimal base station for executing the EDM task, we can define the action space as $A = \{k_1, k_2, k_3, .... k_K\}$. Therefore, the action space of the agent is discrete, where $K$ is the total number of base stations.

\subsection{Reward Function} 

The reward motivates an agent to make the optimal decision (toward the objective) in each state. We consider a pseudo-sparse reward function where an agent gets a reward after executing a single EDM task. In contrast, each state consists of features related to a single smart meter. Hence, the agent determines a specific server, executes the EDM task, and supplies reward to each smart meter's offloading decision that takes part in EDM task execution. We can define immediate reward as $R = -\delta_1 t^j_{EDM} - \delta_2 E^j_{EDM}$, 
where $\delta_1$ and $\delta_2$ are balancing parameters. The reward motivates the agent to select the edge server where the execution cost of the EDM task is lower, considering the adjacent participating smart meters.

\subsection{Deep Q-Learning Method}

Since the state space of the offloading DRL is continuous and extensive, we utilize a deep neural network-based function approximator to limit the agent's training period. In this paper, leveraging the DQN model, a replay buffer is used to store experiences to train the neural network. The stored experiences contain tuples of $<s_t, a_t, r_t, s_{t+1}>$ where $s_t, a_t, r_t$ denotes state, action and reward, at a certain state and $s_{t+1}$ denotes the next state.
We assume that a trusted entity (i.e., mobile switching center (MSC)) will be responsible for training the DRL agent, and the trained policy is fed back to each base station for every tenure. In order to determine the optimal base station at each tenure, the distributed base stations can make decisions independently. We assume that base stations do not manipulate the learned policy during the offloading decision-making process.

\section{FSI Attack and Detection} \label{s:attack}

\subsection{FSI Attack Method}

\begin{figure}[t]
\centerline{\includegraphics[width=0.5\textwidth]{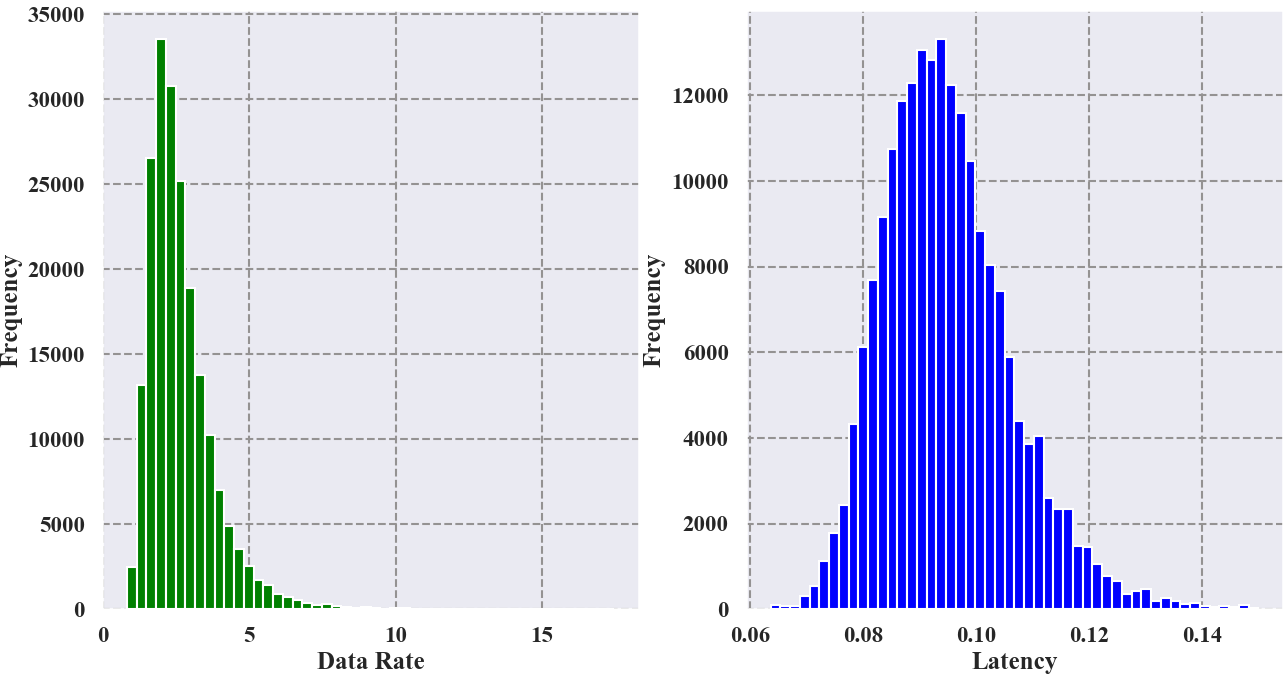}}
\caption{Exploiting data distribution: adversary examining the distribution of data rate and latency of a base station.}
\label{fig:result_0}
\end{figure}

To launch an FSI attack, the attacker is required to perform a reconnaissance of the distribution of state-space features. This paper considers that an attack is launched from the server end, either by the edge servers themselves or by an external entity manipulating the communication channel. Hence, the adversary must select the server-side state features (i.e., data rate or average queuing latency). For simplicity, we consider that the adversary can only conjecture the data rate $(DR)$ distribution for tampering with the state space. Fig. \ref{fig:result_0} illustrates the distribution of data rate for a specific server. Since the distribution is normal, we consider that the attacker can launch the attack by manipulating the mean of the distribution. Therefore, we can define the adversarial attack distribution \cite{giraldo2020adversarial}, by,
    $f^*_a(DR) = \frac{1}{\sqrt{2\pi}\sigma_{DR}} e^{-{\frac{(DR-\theta_{DR}-\sqrt{2\gamma}\sigma_{DR})^2}{2\sigma_{DR}^2}}}$,
 where the expected value of data rate $(\mu_{DR})$ distribution is shifted by, $\mu_{DR} = \theta_{DR} + \sqrt{2\gamma}\sigma_{DR}$,  $\theta_{DR}$ and $\sigma_{DR}$ are the mean and standard deviation of the data rate distribution, and $\gamma$ is the FSI injection control parameter which is controlled by the adversary.\looseness=-1
\subsection{FSI Detection Mechanism} \label{s:detectionMechanism}

To detect false states in the state space, we deploy a state anomaly detection technique on the agent. We utilize lightweight supervised learning models to develop the FSI detection model. We analyze the performance of different classification methods, such as random forest (RF), decision tree (DT), XGBoost, SVM, and neural networks (NN). We assume that a trusted entity will survey the state space and create a labeled dataset to train the models. 
\section{Numerical Analysis} \label{s:numericalAnalysis}
To measure the intensity of an attack, we choose the term ``cost of offloading'', which is the sum of both the latency and the energy consumption. 
\vspace{-5pt}

\subsection{Simulation Setup} 
We consider that four different base stations support a number of smart meters within a region of $400$m$\times400$m. We assume that the bandwidth for each base station is $10$ MHz, and the transmission power consumed by each smart meter is $500mW$, the computation capacity of each base station is $16GHz$. We also assume that the demand and monitoring data sizes are randomly distributed in $[100, 400]Kb$ and $[0.5, 1]Mb$, respectively. Similarly, the CPU cycle requirements for the demand scheduling and forecasting task are randomly distributed in $[0.1, 0.4]GHz$ and $[0.5, 1]GHz$. We assume that four different base stations provide edge computation service to smart meters and the adversary can manipulate the feature space of up to three base stations.

\subsection{Data Distribution Analysis}

We analyze the data distribution for state-space features, such as data rate $(DR)$ and latency $(L)$, and illustrate the results in Fig. \ref{fig:result_0}. In both cases, we observe that the data are normally distributed with different means and standard deviations. Therefore, to launch a stealthy FSI attack, the attacker needs to determine specific features with appropriate values. Otherwise, anomaly detectors have a high probability of detecting the attack. The attacker needs to infer the mean and standard deviation of the data distribution to successfully inject the FSI attack while evading the attack detection system.
\begin{figure*}[!t]
    \centerline{\includegraphics[width=1\linewidth]{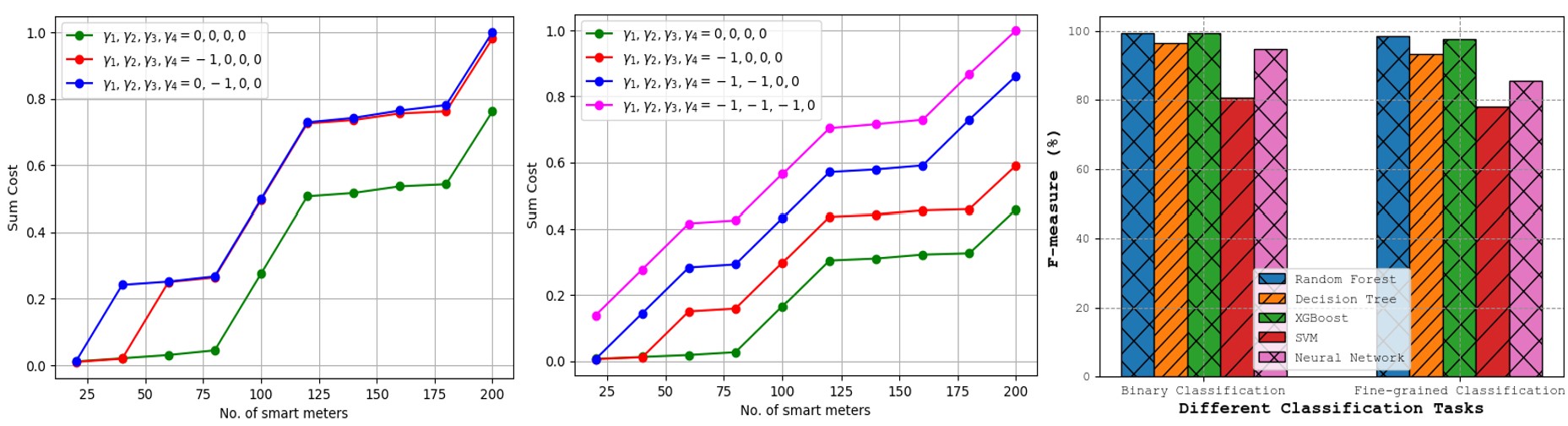}}
    \caption{(a) Analyzing the effect of FSI attack on a single base station, (b) Analyzing the effect of false state injection (FSI) attacks on multiple base stations, (c) Performance of different classification models in detecting FSI attacks.
    }
    \label{fig:ImpactAnalysis}
\end{figure*}
\subsection{False State Injection Attack Impact Measurement}
Figs. $3a$ and $3b$ illustrate the impact of the FSI attack on the offloading cost when the attack is launched on a single base station and multiple base stations. We can infer from the figures that the attack impact is positively correlated with the number of compromised base station features. In Fig. $3a$, we can observe that the impact of compromising any base station is prominent in the increase of the offloading cost. However, other factors, i.e., data rate or latency, could potentially have an impact on such an increment (we leave this analysis as our future endeavor). In Fig. $3b$, we can observe that the offloading cost is doubled if the compromised base station is increased to $2$ from $1$. This observation is also consistent with the increase in compromised base stations from $1$ to $3$, as well. Here, we use the normalized sum cost as the metric.

\subsection{False State Injection Attack Detection}

We analyze the performance of different classification models in FSI attack detection, and the results are illustrated in Fig. $3c$. By examining the results, we can observe that the RF-based classifier attains higher accuracy in detecting FSI attacks both in binary and fine-grained classification. The advantage of fine-grained classification is that it can detect the base station whose features are manipulated. In contrast, the binary ones, after the detection of the anomalous feature, can not provide any additional fine-grained information. To evaluate the performance of the classifier in FSI attack detection, we use the compound F-measure metric. The F-measure metric considers both the recall and the precision of the underlined method. The highest F-measure (of around $99\%$) is achieved by the RF and XGBoost classifiers, while the SVM classifiers produce the lowest score, around $80\%$. 
The remaining classifiers, i.e., DT and NN, attain F-measure values within the $83\%-98\%$ range as presented in Fig. $3c$ .

\section{Conclusions} \label{s:Conclusion}

This paper proposes a DRL-based EDM task offloading scheme for a 6G-enabled SG that features smart meters and harnesses their computational capabilities. According to our results, the impact of FSI attacks on the offloading cost is prevalent and significantly increases the cost with respect to the intensity of the FSI attack. We develop a lightweight supervised FSI detection technique to detect false state estimations on distributed offloading agents. This detection technique can identify a compromised base station with high confidence. Our future work will investigate few shot learning-based offloading techniques, as reinforcement learning requires a considerable number of iterations to converge.


\bibliography{reference}
\nocite{*}
\bibliographystyle{IEEEtran}

\end{document}